\documentclass[runningheads,a4paper]{llncs}

\usepackage[pdftex]{graphicx}

\usepackage[caption=false,font=normalsize,labelfont=sf,textfont=sf]{subfig}

\usepackage{url}
\usepackage{float}
\usepackage{multirow}

\usepackage{array,booktabs,ragged2e}
\newcolumntype{L}[1]{>{\RaggedRight\arraybackslash}p{#1}}
\newcolumntype{R}[1]{>{\RaggedLeft\arraybackslash}p{#1}}

\begin{document}

\title{Discovering Business Area Effects To Process Mining Analysis Using Clustering and Influence Analysis}

\titlerunning{Discovering Business Area Effects To Process Mining Analysis...}

%\author{Omitted, Double Blind}
%\authorrunning{Omitted, Double Blind}
%\institute{Omitted, Double Blind}

\author{Teemu Lehto\inst{1,2}\orcidID{0000-0002-1332-1853} \and Markku Hinkka\inst{1,2}\orcidID{0000-0002-3679-3677} }
\authorrunning{Teemu Lehto, Markku Hinkka}
\institute{
	QPR Software Plc, Finland
	\and
	Aalto University, School of Science, Department of Computer Science, Finland
}

\toctitle{Discovering Business Area Effects to Process Mining Analysis Using Clustering and Influence Analysis}

%\tocauthor{Teemu Lehto, Markku Hinkka, Alexander Jung}
\tocauthor{Teemu Lehto, Markku Hinkka}
\maketitle

\begin{abstract}
A common challenge for improving business processes in large organizations is that business people in charge of the operations are lacking a fact-based understanding of the execution details, process variants, and exceptions taking place in business operations. While existing process mining methodologies can discover these details based on event logs, it is challenging to communicate the process mining findings to business people. In this paper, we present a novel methodology for discovering business areas that have a significant effect on the process execution details. Our method uses clustering to group similar cases based on process flow characteristics and then influence analysis for detecting those business areas that correlate most with the discovered clusters. Our analysis serves as a bridge between BPM people and business, people facilitating the knowledge sharing between these groups. We also present an example analysis based on publicly available real-life purchase order process data.


\end{abstract}

\section{Introduction}

Process mining helps organizations to improve their operations by providing valuable information about the business processes in easy to understand visual flowchart format based on transactional data in ERP systems. However, to provide these meaningful results, the data extracted from ERP systems may often contain different kinds of objects like 'apples and oranges' that should be analyzed separately. Using the procurement process as an example: the purchase order database tables may contain several different kinds of purchase orders items like services, equipment, raw materials, software licenses, high-cost items, free items, headquarter purchases, plant maintenance costs, manually approved items, and automatic replenishment purchases. Without appropriate tools the process analyst needs to either a) analyze all items separately - leading to potentially massive amount of work, b) analyze all items at the same time - leading to potentially meaningless results or c) rely on subjective information like asking business people which of the items should be analyzed separately or relying on own intuition. The techniques presenting in this paper help the analyst to discover those business areas (classification rules) that seem to have a major effect on the business process flow. These business areas are based on case attribute characteristics of the cases and thus easy to understand for the business people. Discovered business areas can be used to effectively guide the process mining analysis further in divide \& conquer manner.

In this paper, we present methods to answer these three questions:
\begin{itemize}
\item How a business process can be analyzed based on the process flow of individual process instances in order to discover business-relevant clusters in such a way that a business analyst can easily understand the clustering results and use them for further analysis. 
\item How to find business areas that have a major effect on process flow behavior.
\item How to further consolidate business area results to discover case attributes that have a significant effect on process flow behavior.
\end{itemize}

The rest of this paper is structured as follows: Section~\ref{relatedwork} is a summary of the latest developments. Section~\ref{Discovering Business Area Effects} present our methodology for Discovering Business Area Effects. Section~\ref{Case Study: Purchase Order process} is a case study with real-life purchase order process data. Section 5 shows limitations and Section~\ref{summary_and_conclusions} draws the final conclusions.

\section{Related Work}
\label{relatedwork}

Process mining is an active research area that analyses business processes based on the event log data from IT systems in order to discover, monitor, and improve processes \cite{aalst_2012}. Process mining typically focuses on discovering the process flowchart as a control flow diagram, Petri net, or BPMN diagram. Other process mining types include conformance checking and enhancement. Root cause analysis as part of process mining has been studied in \cite{suriadi} as well as in our previous works \cite{lehto} and \cite{lehto2}.

One key challenge in process mining is that a single event log may often contain many different processes, in which case trying to discover a single process diagram for the whole log file is not a working solution. In the process mining context clustering has been studied a lot for with excellent results \cite{de_medeiros}, \cite {song}, \cite{de_leoni} and \cite{thaler}. These previous work cover the usage of several distance measures like Euclid, Hamming, Jaccard, Cosine, Markov chain, Edit-Distance as well as several cluster approaches like partitioning, hierarchial, density-based and neuronal network. However, most of the previous research related to clustering within the process mining field has been directly focused on the process flowchart discovery with the prime objectives categorized as Process, Variant or Outlier Identification, Understandability of Complexity, Decomposition, or Hierarhization. In practice, this means that clustering has been used as a tool for improving the other process mining methods like control flow discovery to work better, i.e., clustering has divided the event log into smaller sub logs that have been directly used for further analysis. In this paper, we show how to use clustering for discovering those business areas that have a significant effect on process behavior. Yet another use case for clustering in the process mining field has been to perform structural feature selection in order to improve the prediction accuracy and performance \cite{hinkka}.

Some recent research has started to address the challenge of how to explain the clustering results to business analyst \cite{de_koninck}. It has been presented that when explaining the characteristics of clusters to business analysts, the role of case attributes becomes more important \cite{seeliger}. We show an easy-to-understand representation for showing cluster characteristics based on the difference of densities and case attribute information.

Substantial effort has also been spent in the process mining community to discover branching conditions from business process execution logs \cite{Leoni2}. This has also lead to the introduction of decision models and decision mining \cite{bazhenova} as well as a standard Decision Model and Notation (DMN) \cite{OMG_DMN}. While the objective of the decision modeling is to provide additional details into individual branching conditions, our approach is to analyze the effect of any business area to the whole structure of the process flow, not just one decision branch at a time.

\section{Discovering Business Area Effects}
\label{Discovering Business Area Effects}

In this section, we present our methodology for Discovering Business Area Effects To Process Mining Analysis Using Clustering and Influence Analysis. Our approach is to do the clustering using process flow features and then use influence analysis to find those business areas that have the highest contribution for certain kinds of cases ending up in distinct clusters. If all process instance-specific business area values derived using any given case attribute are distributed randomly, then the contribution measure for each business area is zero, and the information for the analyst is that the particular case attribute does not correlate with the way how the clusters are formed. According to our methodology, it then means that the particular case attribute has no influence on the process flow behavior. In summary, our method finds those business areas and case attributes that have the highest contribution to the process flow behavior.

\subsection{Clustering Cases}

\subsubsection{Feature Selection}

To identify those business areas that have the strongest effect on the process execution, we first run clustering using relevant features representing the process execution characteristics. These features have been widely studied in Trace Clustering papers \cite {song},\cite{thaler}, \cite{hinkka} and \cite{hinkka2}.  Clustering is a trade-off between quality and performance. As the amount of features is increased, the quality of the results potentially improves while performance gets slower.
\begin{itemize}
\item {Activity profile}: This profile contains one feature for each Event Type label in the data. The value of this feature is related to the number of occurrences of that particular event type within the case. If the number of occurrences is used as an exact value, then the clustering algorithm somehow needs to take into account the continuous values, ie. repeating activity A seven times is much more similar to repeating 6 or 8 times, compared to repeating the activity A only twice. One approach is to use value \textit{zero} if the Event Log contains no occurrences of the Event Type for the given case and \textit{one} if the log contains one or more occurrences. While this approach often works well, it may not be able to detect the repeating of a given Event Type multiple times with the log. For this reason, we recommend using value \textit{zero} for no occurrences of the Event Type, \textit{one} for only one occurrence and two for \textit{two} or more occurrences.
\item {Transition profile}
The transition profile captures all process flows from every activity to the next activity. In effect, it contains the process control flow information. Transition profile potentially provides a large number of features up to the square of the number of Event Types plus one for start and end transitions. For example, in the sample analysis presented in Section~\ref{Case Study: Purchase Order process}, we have 42 distinct event types, giving potentially $43^2=1849$ distinct transition. Luckily the control flow for 251.734 cases only contains 676 distinct transitions. Because the amount of transition features is high, we recommend using the coding \textit{zero} if the transition does not occur in the case and \textit{one} if it occurs once or more.

\end{itemize}

\subsubsection{Clustering Algorithms}

A comparative analysis of process instance cluster techniques has been presented in \cite{thaler} and shows how various clustering techniques have been used to separate different process variants from a large set of cases as well as reducing the complexity by grouping similar cases into same clusters. Considering our method, the main functional requirement for the clustering algorithm is that it needs to put cases with similar process flow behavior into the same clusters, and all 20 approaches listed in \cite{thaler} meet this requirement. If a particular clustering algorithm produces meaningful results and if there indeed is a correlation with a particular business area, then our method gives very high contribution values for that business area. If the clustering algorithm does not work perfectly but is still capable to some extent grouping similar cases together, then the contribution values are still likely to show the most significant business areas among the top contributors. The essential non-functional requirement for the clustering algorithm is performance, i.e., the ability to produce results fast with a small amount of memory. With these considerations, we have received good results with the algorithms and parameters below: 

\begin{itemize}
\item \textit{One-hot encoding.} Since our Activity and Transition feature profiles only include categorial values zero, one, and two, it is possible to use efficient one-hot encoding. This results in maximum of ($n(Event Types)+1)^2 + 2 * n(Event Types)$ feature vectors.
\item \textit{Hamming distance} is the natural choice as the distance function with binary data like one-hot encoded features, because it completely avoids the floating-point distance calculations needed for common Euclid distance measure.
\item \textit{K-modes} clustering algorithm is suitable for categorical data. In our tests, k-modes produced well-balanced clusters and was fast to execute. The result of K-modes depends on the initial cluster center initialization. We also tested agglomerative clustering algorithms, but it produced highly unbalanced clusters.
\item \textit{Number of clusters} has a significant effect on the clustering. To discover the business areas, clustering should be done several times with different numbers of clusters. We found out that clustering four times with cluster sizes 2, 3, 5, and 10 clusters gave enough variation in the results providing meaningful results. When the number of clusters is less than five, the large business areas correlate more with the clustering. While clustering to 10 or more clusters, the smaller business areas like \textit{Vendor}, \textit{Customer}, \textit{Product} having more distinct values correlate more with the clusters. Running the clustering several times is also an easy way to mitigate the random behavior of K-modes coming from initialization. 
\end{itemize}

\subsection{Influence Analysis}
\label{Detecting changes using Influence Analysis}

\subsubsection{Business Areas} Examples of business area dimensions include: \textit{company code}, \textit{product line}, \textit{sales unit}, \textit{delivery team}, \textit{geographical location}, \textit{customer group}, \textit{product group}, \textit{branch offices}, \textit{request category} and \textit{diagnosis code}. All the case attributes that are relevant to business can be used as business area dimensions as such, for example, \textit{product code}. However, a large organization may easily have thousands of low-level product codes in their ERP system, so it is beneficial to have access to product hierarchy and use each level as a separate business area dimension. Another example of a derived business area dimension is when a case attribute like \textit{Logistics Manager Name} can be used to identify the \textit{Delivery Team}. We again suggest having both the \textit{Logistics Manager} and \textit{Delivery Team} as business area dimensions; if one particular \textit{Logistics Manager} has many cases and a major effect on process flow behavior then our method will show that person as the most significant business area in the \textit{Logistics Manager} dimension. The third example of derived business areas is to utilize the event attributes. For example the \textit{Logistics Manager Name} may be stored as an attribute value for the \textit{Delivery Planning Done} activity. If there is always at most one \textit{Delivery Planning Done} activity, then the attribute value can be used as such in the case level. If there are multiple \textit{Delivery Planning Done} activities, then typical options include: use the first occurrence, use the last occurrence or use a concatenated comma-separated list of all distinct values from activities as the value on the case level. The outcome of forming business area dimensions is a list of case-level attributes that contain a specific (possibly empty) business area value for each case. To continue with our formal methodology, we now consider these business area dimensions as case attributes and the case attribute values as the corresponding business areas.

% One core concept of our methodology is to analyze the relationship of the process flow features included in clustering and the business area information to be used in influence analysis. The Influence Analysis has been presented in \cite{lehto} and \cite{lehto2} for analyzing root causes for process mining findings based on case attribute values and derived category dimensions. Forming the case specific business areas can by done in fowwloi relevant businss Starting point for detecting business areas is the process mining attribute profile, ie. the case attribute and event attribute values. Our
  
%Examples of business area are: company codes, product lines, sales units, delivery units, geographical locations, customer groups, product groups, branch offices. Detailed level business areas include the lowest level granularity available in ERP systems, for example individual customer names, delivery addresses, product codes, sales person names and project manager names. Business areas can also be calcuated and formed based on other info. Low level info, grouping and combination of several other business areas.  

\subsubsection{Interestingness Measures}

We now present the definitions for interestingness measures used for finding the business areas that correlate with the clustering results. Let $C = \{c_1, \ldots, c_N\}$ be the set of cases in the process analysis. Each case represents a single business process execution instance. Let $P = \{p_1, \ldots, p_N\}$ be a set of clusters each formed by clustering the cases in $C$. $C_p = \{c_{p_1}, \ldots, c_{p_N}\}$ is the set of cases belonging to cluster p. $C_p \subseteq C$. Similarly $C_a = \{c_{a_1}, \ldots, c_{a_N}\}$ is the set of cases belonging to the same business area a, ie. they have the same value for the case attribute a.

%Using thse definitions we are present the definitions for detecting the characteristics:
%For the purpose of this paper we consider the casea attribute values as discrete values. $C_a \subseteq C$. 

\begin{definition}
Let Density $\rho(a, C) = \frac{n(C_a)}{n(C)}$ where $n(C_a)$ is the total amount of cases belonging to the business area a and $n(C)$ is the total amount of all cases in the whole process analysis. 
Similarly, the Density $\rho(a, C_p) = \frac{n(C_p \cap C_a)}{n(C_p)}$ is the density of cases belonging to the business area a within the cluster P. 
\end{definition}

\begin{definition}
Let $Contribution\%(a \to p) = \rho(a, C_p) - \rho(a, C) =$ \\  $ \frac{n(C_p \cap C_a)}{n(C_p)} - \frac{n(C_a)}{n(C)} $ is the extra density of cases belonging to the business area a in the cluster p compared to average density.
\end{definition}

 If business area a is equally distributed to all clusters, then the \\ $Contribution\%(a\to p)$ is close to zero in each cluster. If the business area a is a typical property in a particular cluster $p_i$ and rare property in other clusters, then the $Contribution\%(a\to p_i)$ is positive and other $Contribution\%(a \to p_j, where j <> i)$ values are negative. Calculating the sum of all Contribution values for all clusters is always zero, so the extra density in some clusters is always balanced by the smaller than average density in other clusters. 

 We now want to find the business areas that have a high contribution in many clustering. We define:
 
% \begin{definition}
% 	Let $ClusterContribution(a \to p) = \frac{n(CL_p)}{n(C)}(Contribution\%(a \to p))^2$.   = extra density of cases belonging to the business area a in the cluster p compared to average density.
% \end{definition}

 \begin{definition}\label{def:BusinessAreaContribution}
	Let $BusinessAreaContribution(a) =$ \\ $\sum\limits_{{p_i} \in P}\frac{n(C_{p_i})}{n(C)}(max\{Contribution\%(a \to p_i),0\})^2$. \\ Here we sum the weighted squares of all positive contributions the business area a has with any clustering $p_i$. Positive values of $Contribution\%(a \to p_i)$ indicate a positive correlation with the business are a and the particular cluster i, while negative values indicate that the business area a has smaller than the average density in the cluster i. We found out that using only the positive correlations gives more meaningful results when consolidating to the business area level. Since a few high contributions are relatively more important than many small contributions, we use the Variance of the density differences, i.e., taking the square of the $Contribution\%(a \to p_i)$. Since a contribution within a small cluster is less important than contribution in a large cluster, we also use the cluster size based weight $\frac{n(C_p)}{n(C)}$.
 \end{definition}

  Any particular business area a may have a substantial contribution in some clusters and small contribution in other, so the sum of all these clusterings is giving the overall correlation between business area a and all clusters $p_i \subseteq P$

  We use the term \textit{Business area} in this paper for any combination of a process mining case attribute and a distinct value for that particular case attribute. $BusinessAreaContribution$ thus identifies the individual case attribute-value combinations that have the highest effect on clustering results. It is then also possible to continue and consolidate the results further to Case Attribute level:

\begin{definition}
	Let $AT = \{at_1, \ldots, at_N\}$ be a set of case attributes in the process analysis. Each case $c_i \in C$ has a value ${at_j}_{c_i}$ for each case attribute $at_j \in AT$. ${at_j}_{c_i}$  is the value of case attribute $at_j$ for case $c_i$ and $V_{at_j} = \{v_{{at_j}_1}, \ldots, v_{{at_j}_N}\}$ is the set of distinct values that the case attribute ${at_j}$ has in the process analysis. 
\end{definition}

\begin{definition}\label{def:CaseAttributeContribution}
	Let $CaseAttributeContribution(at)$ be a sum of all BusinessAreaContributions from all the business areas corresponding to the given case attribute $at$ as $\sum\limits_{v_{{at_j}_i} \in V_{at_j}} BusinessAreaContribution(at_{j_{v_{{at_j}_i}}})$
\end{definition}

%\begin{definition}
%	Let  $Attribute Contribution_{at_j} =$ \\
%	$Variance[contribution\%(CaseAttributeSubgrouping_{at_j})] =$ \\
%	\\
%	$\sum\limits_{{c_j} \in CaseAttributeSubgrouping_{at_j}} (contribution\%_{c_j}-0)^2 .$
%\end{definition}

\section{Case Study: Purchase Order Process}
\label{Case Study: Purchase Order process}

In this section, we apply our method to the real-life purchase order process data from a large Netherlands multinational company operating in the area of coatings and paints. The data is publicly available as the BPI Challenge 2019 \cite{dongen} dataset. We made the following choices:
\begin{itemize}
\item \textbf{Source data} We imported the data from the XES file as such without any modifications. To keep the execution times short, we experimented with the effect of running the analysis with a sample of the full dataset. Our experiments showed that the results remained consistent for sample size 10.000 cases and more. With the sample size of 1.000 cases, the results of the individual analysis runs started to change, so we decided to keep the sample size 10.000 cases.
\item \textbf{Clustering algorithm} We used the k-modes clustering as implemented in Accord.Net Machine Learning Framework \cite{souza} with one-hot encoding and hamming distance function. To take into account the different clustering sizes, we performed clustering four times, fixed to two, three, five, and ten clusters.
\item \textbf{Activity profile features for clustering} We used our default boolean activity profile, which creates one feature dimension for each activity and the value is \textit{zero} if the activity does not occur in the case, value \textit{one} if the activity occurs once and value \textit{two} if it is repeated multiple times. There were 37 different activities in the sample, and the Top 20 activity profile is shown in Table \ref{table:top20activityprofile}.

\begin{table}[!htbp]
	\caption{Activity profile: Top 20 activities ordered by unique occurrence count}
	\centering
	\begin{scriptsize}
		
	\begin{tabular}{ p{7.0cm} R{2.5cm} R{2.0cm} }
		\toprule
		\multicolumn{1}{l}{\textbf{Name}} & \textbf{Unique Count} & \textbf{Count} \\ 
		\midrule
			Create Purchase Order Item & 10 000 & 10 000 \\ 
			Record Goods Receipt & 9 333 & 13 264 \\ 
			Record Invoice Receipt & 8 370 & 9 214 \\ 
			Vendor creates invoice & 8 310 & 8 901 \\ 
			Clear invoice & 7 245 & 7 704 \\ 
			Remove Payment Block & 2 223 & 2 272 \\ 
			Create Purchase Requisition Item & 1 901 & 1 901 \\ 
			Receive Order Confirmation & 1 321 & 1 321 \\ 
			Change Quantity & 707 & 853 \\ 
			Change Price & 443 & 498 \\ 
			Delete Purchase Order Item & 338 & 339 \\ 
			Cancel Invoice Receipt & 251 & 271 \\ 
			Vendor creates debit memo & 244 & 253 \\ 
			Record Service Entry Sheet & 232 & 10 326 \\ 
			Change Approval for Purchase Order & 194 & 319 \\ 
			Change Delivery Indicator & 112 & 128 \\ 
			Cancel Goods Receipt & 109 & 136 \\ 
			SRM: In Transfer to Execution Syst. & 42 & 57 \\ 
			SRM: Awaiting Approval & 42 & 50 \\ 
			SRM: Complete & 42 & 50 \\ 
						
		\bottomrule
	\end{tabular}
	\end{scriptsize}

	\label{table:top20activityprofile}
\end{table}

\item \textbf{Transition profile features for clustering} Using a typical process mining analysis to discover the process flow diagram, we discovered 376 different direct transitions, including 13 starting activities, 22 ending activities, and 341 direct transitions between two unique activities. All of these 376 features were used as dimensions for clustering in a similar way as the activity profile, i.e., boolean value \textit{zero} if transition did not occur in the case and \textit{one} if it occurred once or multiple times.

\item \textbf{Business area dimensions} Since we did not have any additional information or hierarchy tables concerning possible business areas, we are using all available 15 distinct case attributes listed in Table \ref{table:caseattributecontribution} as business area dimensions. These case attributes have a total of 9901 distinct values, giving us 9901 business areas to consider when finding those business areas that have the most significant effect on process flow. 

\end{itemize}

\subsection{Clustering Results for Individual Clustering}
\label{Clustering results}

Table \ref{table:ResultsByContribution} shows the results of clustering to fixed five clusters. We see that the first cluster contains 48\% of cases, the second cluster 33\%, third 17\%, and both 4th and 5th one percent each. Here we show the five most important business areas based on the contribution\%, which is calculated as the difference between Cluster specific density of that business area and Total Density. These results already give hints about the meaningful characteristics in the whole dataset, ie: Cluster one contains many \textit{Standard} cases from spend areas related to \textit{Sales}, \textit{Products for Resale} and \textit{NPR}. On the other hand cluster two contains more than average amount of cases from \textit{spend area} \textit{Packaging}, related to \textit{Labels} and \textit{PR}. \textit{VendorID\_0120} seems to be highly associated with the process flow characteristics of cluster 2. Cluster 3 is dominated by \textit{Consignment} cases. Cluster 4 contains many  \textit{Metal Containers \& Lids} cases as well as cases from \textit{VendorID}s \textit{0404} and \textit{0104}. Further analysis of the top five business areas listed as characteristics for each cluster confirms that these business areas indeed give a good overall idea of the cases allocated into each cluster.

\begin{table}[ht]
	\caption{Clustering results based on Contribution}
	\centering
\begin{scriptsize}
	\begin{tabular}{ p{1.3cm} p{6.0cm} R{1.2cm} R{1.2cm} R{1.9cm} }
		\toprule
		\textbf{Cluster} & \multicolumn{1}{l}{\textbf{Business Area $a$}} & \textbf{Cluster Density} & \textbf{Total Density} & \textbf{Contribution}  \\ 
		\midrule
 			& Spend area text = Sales & 0.36 & 0.26 & 0.11 \\ 
Cluster1 	& Sub spend area text = Products for Resale & 0.34 & 0.24 & 0.11 \\ 
48\% cases 	& Spend classification text = NPR & 0.41 & 0.32 & 0.10 \\ 
			& Item Type = Standard & 0.96 & 0.87 & 0.09 \\ 
			& Item Category = 3-way match, invoice before GR & 0.95 & 0.88 & 0.07 \\
		\hline
 			& Spend area text = Packaging & 0.65 & 0.44 & 0.21 \\ 
Cluster2	& Sub spend area text = Labels & 0.39 & 0.24 & 0.16 \\ 
33\% cases	& Spend classification text = PR & 0.79 & 0.66 & 0.13 \\ 
			& Name = vendor\_0119 & 0.14 & 0.05 & 0.08 \\ 
			& Vendor = vendorID\_0120 & 0.14 & 0.05 & 0.08 \\ 
		\hline
			& Item Category = Consignment & 0.33 & 0.06 & 0.27 \\ 
Cluster3	& Item Type = Consignment & 0.33 & 0.06 & 0.27 \\ 
17\% cases	& Name = vendor\_0185 & 0.09 & 0.02 & 0.08 \\ 
			& Vendor = vendorID\_0188 & 0.09 & 0.02 & 0.08 \\ 
			& Item = 10 & 0.33 & 0.26 & 0.07 \\ 
		\hline
			& Sub spend area text = Metal Containers \& Lids & 0.19 & 0.08 & 0.11 \\ 
Cluster4 	& Name = vendor\_0393 & 0.09 & 0.01 & 0.08 \\ 
1\% cases   & Vendor = vendorID\_0404 & 0.09 & 0.01 & 0.08 \\ 
			& Name = vendor\_0104 & 0.11 & 0.04 & 0.07 \\ 
			& Vendor = vendorID\_0104 & 0.11 & 0.04 & 0.07 \\ 
		\hline
			& Spend classification text = NPR & 0.59 & 0.32 & 0.27 \\ 
Cluster5	& Spend area text = Sales & 0.41 & 0.26 & 0.15 \\ 
1\% cases   & GR-Based Inv. Verif. = TRUE & 0.21 & 0.06 & 0.15 \\ 
			& Item Category = 3-way match, invoice after GR & 0.21 & 0.06 & 0.15 \\
			& Sub spend area text = Products for Resale & 0.38 & 0.24 & 0.14 \\ 
	
		\bottomrule
	\end{tabular}
\end{scriptsize}

	\label{table:ResultsByContribution}
\end{table}

\subsection{Discovering Business Areas}
\label{Clustering summary for Business Areas}

We clustered four times for fixed cluster amounts of 2,3,5 and 10 - yielding a total of 20 clusters, and then consolidating the results into business area level using Definition \ref{def:BusinessAreaContribution}. The top 20 of all these 9901 business areas ordered by their respective Business Area Contribution is shown in Table \ref{table:businessareacontribution}. Clearly the business areas \textit{Item Category = Consignment} and \textit{Item Type = Consignment} have most significant effect on the process flow. Looking at the actual process model, we see that \textit{Consignment} cases completely avoid three of the five most common activities in the process, namely \textit{Record Invoice Receipt}, \textit{Vendor creates invoice} and \textit{Clear Invoice}.
Similarly, the business area \textit{Spend area text = Packaging} also has a high correlation with process flow characteristics. Analysis of the process model shows that, for example, 23\% of \textit{Packaging} cases contain activity \textit{Receive Order Confirmation} compared to only 5\% of the other cases. Further analysis of all the business areas listed in Table \ref{table:businessareacontribution} shows that each of these areas has some distinctive process flow behavior that is more common in that area compared to the other business areas.

\begin{table}[!htbp]
	\caption{Top 20 Business areas with major effect to process flow}
	\centering
	\begin{scriptsize}
	
	\begin{tabular}{ p{7.0cm} R{2.2cm} R{2.4cm} }
		\toprule
		\multicolumn{1}{l}{\textbf{Business Area $a$}} & \textbf{Contribution} & \textbf{nCases $n(C_a)$} \\ 
		\midrule

Item Category = Consignment & 0.051 & 576 \\ 
Item Type = Consignment & 0.051 & 576 \\ 
Spend area text = Packaging & 0.040 & 4382 \\ 
Spend classification text = NPR & 0.024 & 3175 \\ 
Sub spend area text = Labels & 0.022 & 2351 \\ 
Spend area text = Sales & 0.021 & 2574 \\ 
Item Type = Standard & 0.021 & 8740 \\ 
Sub spend area text = Products for Resale & 0.021 & 2390 \\ 
Spend classification text = PR & 0.019 & 6574 \\ 
Item Category = 3-way match, invoice before GR & 0.017 & 8760 \\ 
Spend area text = Logistics & 0.013 & 210 \\ 
Item Type = Service & 0.013 & 244 \\ 
Item = 1 & 0.012 & 342 \\ 
GR-Based Inv. Verif. = TRUE & 0.012 & 623 \\ 
Item Category = 3-way match, invoice after GR & 0.012 & 625 \\ 
Name = vendor\_0119 & 0.007 & 549 \\ 
Vendor = vendorID\_0120 & 0.007 & 549 \\ 
Sub spend area text = Road Packed & 0.006 & 145 \\ 
Name = vendor\_0185 & 0.004 & 163 \\ 
Vendor = vendorID\_0188 & 0.004 & 163 \\ 
		
		\bottomrule
	\end{tabular}
\end{scriptsize}
	\label{table:businessareacontribution}
\end{table}

\subsection{Clustering Summary for Case Attributes}
\label{Clustering summary for Case Attributes}

Finally, Table \ref{table:caseattributecontribution} consolidates individual business areas into the Case Attribute level. \textit{Item Type} with six distinct values and \textit{Item Category} with four distinct values have most significant effects on process flow characteristics. To confirm the validity of these results we further analysis the materials provided in BPI Challenge 2019 website including the background information and submission reports \cite{dongen}. It is clear that the \textit{Item Type} and \textit{Item Category} indeed can be regarded as the most important factors explaining the process flow behavior as they are specifically mentioned to \textit{roughly divide the cases into four types of flows in the data}. It is also interesting to see that the \textit{Spend are text} and \textit{Sub spend are text} have a significant effect on the process flow even though they have much higher number of distinct values (19 and 115) compared to \textit{Spend classification text} which only has four distinct values.  

\begin{table}[!htbp]
	\caption{Case Attributes ordered by effect on process flow}
	\centering
	\begin{scriptsize}
		
	\begin{tabular}{ p{4.0cm} R{2.2cm} R{3.9cm} }
		\toprule
		\multicolumn{1}{l}{\textbf{Case Attribute $at$}} & \textbf{Contribution} & \textbf{Distinct Values $n(V_{at})$} \\ 
		\midrule
		
		Item Type & 0.086 & 6 \\ 
		Item Category & 0.080 & 4 \\ 
		Spend area text & 0.077 & 19 \\ 
		Sub spend area text & 0.056 & 115 \\ 
		Spend classification text & 0.043 & 4 \\ 
		Name & 0.025 & 798 \\ 
		Vendor & 0.025 & 840 \\ 
		Item & 0.016 & 167 \\ 
		GR-Based Inv. Verif. & 0.012 & 2 \\ 
		Purchasing Document & 0.002 & 7937 \\ 
		Document Type & 0.000 & 3 \\ 
		Goods Receipt & 0.000 & 2 \\ 
		Company & 0.000 & 2 \\ 
		Source & 0.000 & 1 \\ 
		Purch. Doc. Category name & 0.000 & 1 \\ 
		
		\bottomrule
	\end{tabular}
	\end{scriptsize}

	\label{table:caseattributecontribution}
\end{table}

\section{Limitations}
\label{Limitations}
Forming business area dimensions is an essential step in our method. However, some relevant business areas may consist of several dimensions, for example, the process flow behavior could be very distinctive in a particular combination of business areas \textit{SalesOffice=Spain} and \textit{ProductGroup=Computers}. Automatically detecting this kind of significant combined business areas would be a useful feature.
Another limitation is that the process flow behavior does not take into account the performance profile, i.e., the lead times between individual activities and the total case duration. Although the usage of this kind of numerical information would require a more advanced clustering technique, the influence analysis part of the method presented in this paper would already handle the discovery of related business areas.

\section{Summary and Conclusions}
\label{summary_and_conclusions}

In this paper, we have presented a method for discovering those business areas that have a significant effect on process flow behavior based on clustering and influence analysis. As a summary of our findings:

\begin{itemize}

%\item We emphasize the idea that business people are familiar with the case attribute related information of the individual process instances and process analyst people are more focused on the process flow behavior of individual cases. This finding is based on our experiences from several hundreds of process mining projects where business people are traditionally following reports and dashboards based on case attributes, and process mining experts are enthusiastic about the new process flow based findings. Based on the results presented in this paper we suggest to first identify those business areas having the biggest effect on process flow using out method and then analyze the identified business areas separately from other cases to provide meaningful and easy-to-understand results about the process flow.
\item Our presented method is capable of discovering those business areas that have the most significant effect on the process execution. Our method provides valuable information to business people who are very familiar with case attributes and attribute values but not so familiar with the often technical event type names extracted from transactional system log files. 
\item Our method supports any available trace clustering method. Our case study shows that using the k-modes clustering algorithm with activity and transition profiles provides good results. 
\item Clustering makes the analysts realize that not all the cases in the process model are similar. Using the \textit{Contribution\%} measure to explain clustering results works well for explaining the clustering results to business people.
\item The case study presented in this paper confirms that the identified business areas indeed have distinctive process flow behavior, for example missing activities, higher than average amount of some special activities, or distinctive execution sequence for activities. Using our method, the business analyst may now divide the process model into smaller subsets and analyze them separately. It is a good idea to start the analysis of any process subset again by running the clustering to see if the cases are similar enough from both process flow point of view.
\item Clustering reduces the need for external subject matter business experts. Naturally, it would be nice to have a person who can explain everything, but in real life, those persons are very busy, and some important details are always likely to be forgotten by busy business people.
\end{itemize}

\subsubsection{Acknowledgements.}  

We thank QPR Software Plc for the practical experiences from a wide variety of customer cases and for funding our research. The algorithms presented in this paper have been implemented in a commercial process mining tool QPR ProcessAnalyzer.

%\section*{Acknowledgment}
%We want to thank QPR Software Plc for funding our research.



\bibliographystyle{IEEEtran}

%%\bibliography{paper}

\end{document}